\documentclass{article}

\usepackage{arxiv}

\usepackage[utf8]{inputenc} % allow utf-8 input
\usepackage[T1]{fontenc}    % use 8-bit T1 fonts
\usepackage{url}            % simple URL typesetting
\usepackage{booktabs}       % professional-quality tables
\usepackage{amsfonts}       % blackboard math symbols
\usepackage{nicefrac}       % compact symbols for 1/2, etc.
\usepackage{microtype}      % microtypography
\usepackage{lipsum}
\usepackage{graphicx}
\graphicspath{ {./images/} }

\usepackage[T1]{fontenc}
\usepackage{graphicx}
\usepackage{amsmath,amssymb}
\usepackage{bm}
\usepackage{cite}
\usepackage{booktabs}
\usepackage{array}
\usepackage{multirow}
\usepackage{makecell}
\usepackage{siunitx}
\usepackage{xcolor}
\usepackage[hidelinks,breaklinks=true]{hyperref}
\usepackage{url}
\usepackage{tikz}
\usetikzlibrary{arrows.meta, positioning, fit, calc, shapes.geometric}

\DeclareMathOperator{\relu}{relu}

\newcommand{\dn}{\ensuremath{\downarrow}}
\newcommand{\up}{\ensuremath{\uparrow}}

\newcommand{\Lone}{\ensuremath{\mathcal{L}_{1}}}
\newcommand{\Llpips}{\ensuremath{\mathcal{L}_{\mathrm{LPIPS}}}}
\newcommand{\Lunder}{\ensuremath{\mathcal{L}_{\mathrm{under}}}}
\newcommand{\Lseg}{\ensuremath{\mathcal{L}_{\mathrm{seg}}}}
\newcommand{\Ldseg}{\ensuremath{\mathcal{L}_{\mathrm{dseg}}}}

\usepackage{enumitem}   

\title{	MIRAGE: Multi-scale Lesion-Informed Representation with Auxiliary Guidance for MRI Contrast Enhancement
}

\newcommand{\shorttitle}{MIRAGE: Multi-scale Lesion-Informed Representation with Auxiliary Guidance for MRI Contrast Enhancement}
\newcommand{\shortauthors}{Andrea Borghesi et al.}

\author{
 Andrea Borghesi \\
  Netherlands Cancer Institute\\
   \And
 Xin Wang\\
  Netherlands Cancer Institute\\
  \And
 Jonas Teuwen \\
  Netherlands Cancer Institute\\
  \texttt{j.teuwen@nki.nl}
  \And
  George Yiasemis \\
    Netherlands Cancer Institute\\
}

\begin{document}
\maketitle

\begin{abstract}
Inferring contrast enhancement from one pre-contrast breast MRI slice is underdetermined: post-contrast appearance contains physiological information that is not uniquely encoded in baseline anatomy. Optimizing only paired pixel fidelity can suppress uncertain lesion enhancement, whereas adversarial or stochastic generative objectives can favor realistic post-contrast appearance without guaranteeing patient-specific lesion fidelity. We introduce MIRAGE, a residual 2D U-Net that combines global reconstruction and perceptual losses with three forms of lesion-aware supervision available only during training: an asymmetric penalty for missed tumor enhancement, multi-scale auxiliary tumor segmentation, and guidance through a frozen post-contrast tumor segmentation nnU-Net. We evaluate the method on 301 cases from the multi-centre MAMA-SYNTH data using eight complementary image-, region-, radiomics-, and segmentation-based metrics. MIRAGE ranks first on six metrics and markedly improves downstream lesion localization over tuned pix2pix, conditional diffusion, and latent bridge-matching baselines. The generative alternatives retain advantages in LPIPS or contrast classification, revealing a clear fidelity-utility trade-off. Leave-one-in and leave-one-out ablations show that the losses are partly redundant for lesion localization but exert distinct effects on appearance, radiomics, and boundary accuracy. These results support task-aware synthesis while also showing that its apparent optimality is conditional on the downstream models and metrics used to define utility.
 
\keywords{Breast MRI \and DCE MRI \and Virtual contrast enhancement \and Image synthesis \and Task-aware learning \and Deep supervision}
% Authors must provide keywords and are not allowed to remove this Keyword section.

\end{abstract}
\section{Introduction}
\label{sec:intro}

Dynamic contrast-enhanced (DCE) MRI is the most sensitive breast-imaging modality and supports lesion detection, disease-extent assessment, treatment planning, and response monitoring~\cite{mann2019breast}. Its value derives from signal changes after administration of a gadolinium-based contrast agent (GBCA). Although contemporary GBCAs have a strong safety record, contrast administration still requires intravenous access, adds time and complexity, is restricted in selected patients, and contributes to environmental contamination~\cite{starekova2024update,dekker2024review}. These considerations motivate contrast-reduced or contrast-free protocols, provided that clinically relevant enhancement information can be recovered reliably.

Virtual contrast enhancement predicts a post-contrast image $y$ from non-contrast input $x$. Because enhancement depends on physiological factors not fully determined by pre-contrast anatomy, multiple targets may be compatible with the same input, particularly when only a single T1-weighted slice is available. Methods must therefore balance patient-specific fidelity with realistic post-contrast appearance, reflecting the perception--distortion trade-off~\cite{blau2018perception}. In medical imaging, however, plausible synthesis may still add, shift, or suppress clinically relevant features~\cite{cohen2018distribution}.

Breast virtual-contrast studies have used multi-sequence inputs, simulated low-dose acquisitions, diffusion-weighted MRI, lesion-weighted adversarial objectives, and downstream segmentation guidance~\cite{chung2022deep,muller2023using,zhang2023synthesis,kim2022tumor,osuala2025gan}. Conditional diffusion has also shown that subtraction targets and tumor-aware losses improve lesion fidelity, while mask conditioning further helps but requires lesion localization at inference~\cite{ibarra2025subddpm}. These results suggest that global pixel or perceptual criteria alone are insufficient. Existing synthesis families make different compromises: pix2pix uses adversarial learning to encourage sharp target-domain appearance~\cite{isola2017pix2pix}; denoising diffusion models represent flexible conditional distributions through iterative sampling~\cite{ho2020ddpm}; and latent bridge matching (LBM) enables one-step translation in latent space~\cite{chadebec2025lbm}. Direct paired regression is more tightly anchored to the observed anatomy but may average uncertain detail.

We investigate this tension under the MAMA-SYNTH protocol~\cite{joship2026mamasynth}, which maps a single fat-suppressed pre-contrast slice to its peak-enhancement counterpart and evaluates global fidelity, tumor-region similarity, radiomic classification, and segmentation by a fixed post-contrast nnU-Net. We propose \textbf{MIRAGE} (\textbf{M}ulti-scale Lesion-\textbf{I}nformed \textbf{R}epresentation with \textbf{A}uxiliary \textbf{G}uidance for Contrast \textbf{E}nhancement), a residual U-Net trained with image-, lesion-, and task-aware supervision. tumor masks guide training but are not supplied at inference. Our contributions are as follows:

\begin{enumerate}[label=(\roman*),nolistsep, labelwidth=!, labelindent=0pt]
    \item A lesion-informed objective combining asymmetric enhancement preservation, multi-scale segmentation, and frozen downstream-model guidance; 
    \item Complementary leave-one-in and leave-one-out analyses that distinguish isolated benefit from redundancy in the full objective; and 
    \item A comparison with adversarial, diffusion, and latent-bridge alternatives, exposing the trade-off between appearance realism and lesion utility.
\end{enumerate}

\section{Materials and Methods}
\subsection{Data and task}
\label{sec:data}

We use the MAMA-SYNTH release derived from MAMA-MIA~\cite{garrucho2025mamamia,joship2026mamasynth}. It contains 1,506 pretreatment breast DCE-MRI examinations from four public cohorts (Duke \cite{saha2021dbc}, ISPY1 \cite{newitt2016ispy1}, ISPY2 \cite{li2022ispy2}, NACT \cite{newitt2016breastnact}), acquired at 1.5 or 3.0\,T across multiple vendors. For each examination, preprocessing selects the post-contrast phase with the highest mean signal inside the annotated tumor and the 2D slice with the largest tumor area. Each case comprises a fat-suppressed pre-contrast image $x$, its peak-enhancement target $y$, and a binary tumor mask $m$. A global mean and standard deviation estimated from training pre-contrast images are applied to both $x$ and $y$, preserving their relative intensity change. We use 1,205 cases for training and a fixed 301-case internal validation cohort. Masks are never provided to the synthesizer at inference.

\subsection{Architecture and objective}
\label{sec:arch}
% MIRAGE architecture figure.
\begin{figure}[t]
  \centering  \includegraphics[width=\linewidth]{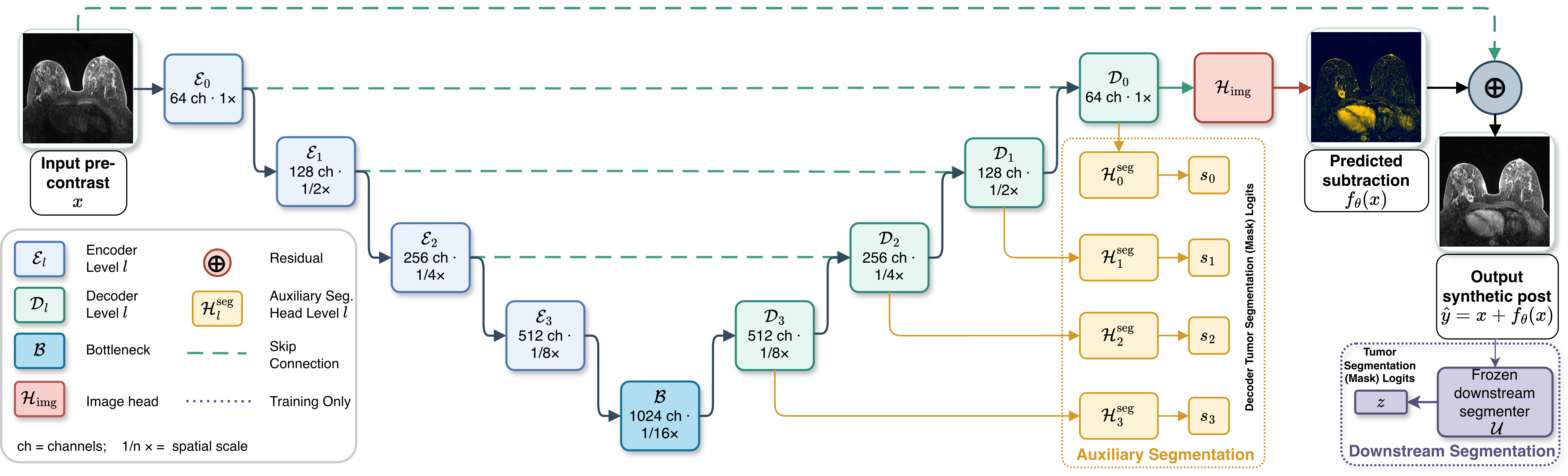}
  \caption{MIRAGE architecture.
    A four-scale U-Net with bottleneck $\mathcal{B}$ and skip connections $\mathcal{E}_l \to \mathcal{D}_l$  maps pre-contrast ${x}$ to synthetic post
    $\hat{{y}}={x}+f_\theta({x})$, where the image head $\mathcal{H}_{\mathrm{img}}$
    predicts the subtraction enhancement $f_\theta({x})$ and a residual connection adds it back to
    ${x}$.
    At inference, only $\hat{{y}}$ is returned.
    During training, segmentation heads on each decoder scale produce tumour-mask
    logits $s_\ell$, supervised against the ground-truth ROI mask ${m}$, and a
    frozen challenge nnU-Net $\mathcal{U}$ segments $\hat{{y}}$ as
    $z$ (after z-normalization) using the same mask for supervision.
    These segmentation paths are outlined with dotted lines and are discarded at
    inference; ${m}$ never enters the synthesiser input.}
  \label{fig:architecture}
\end{figure}

MIRAGE is a four-scale residual 2D U-Net~\cite{ronneberger2015unet} with encoder/decoder widths $[64,128,256,512]$, a $1024$-channel bottleneck $\mathcal{B}$, InstanceNorm, and LeakyReLU. Rather than predicting $y$ directly, an $1{\times}1$ image head $\mathcal{H}_{\mathrm{img}}$ on the finest decoder stage $\mathcal{D}_0$ reads out the predicted subtraction $f_\theta(x)$ and
\begin{equation}
  \label{eq:residual}
  \hat{y} = x + f_\theta(x),
\end{equation}
\noindent
with skip connections from each encoder $\mathcal{E}_l$ to $\mathcal{D}_l$, $l{=}0,\ldots,3$ ($l{=}0$ full res.).

For training-only supervision, four $1{\times}1$ segmentation heads $\mathcal{H}^{\mathrm{seg}}_l$ attach to decoder stage $\mathcal{D}_l$, $l{=}0,\ldots,3$, producing tumor-mask logits at each scale $s_l$. A frozen downstream segmenter (see below) is applied likewise only during training. At inference the network receives only $x$ and returns $\hat{y}$; $\mathcal{H}^{\mathrm{seg}}_l$ and the segmenter path are discarded.
The full architecture is shown in Figure ~\ref{fig:architecture}.

\noindent
The full objective is
\begin{equation}
\label{eq:loss}
    \mathcal{L} = \Lone + 0.3\,\Llpips + 0.05\,\Lunder + 0.1\,\Lseg + 0.2\,\Ldseg ,
\end{equation}
grouped as fidelity, ROI realism, and segmentation. The coefficients were chosen through coarse development sweeps and define one operating point, not a universal optimum. With valid-pixel mask $v$,

\begin{equation}
    \Lone= \frac{\lVert v\odot(\hat y-y)\rVert_1}{\lVert v\rVert_1}.
\end{equation} 

$\Llpips$ is the VGG-based perceptual loss~\cite{zhang2018lpips} on the reconstructed post-contrast image.

To penalise insufficient enhancement in the lesion, define $e=\relu(y-x)$ and $\hat e=\relu(\hat y-x)$. The asymmetric term

\begin{equation}
    \mathcal{L}_{\mathrm{under}}=\frac{\lVert m\odot v\odot\relu(e-\hat e)\rVert_1}{\lVert m\odot v\rVert_1}
    \label{eq:under}
\end{equation}

penalises only ROI pixels where $\hat e<e$; over-enhancement inside the tumor is not penalised by this term and is handled by the global reconstruction losses.

For internal deep supervision, decoder head $l$ predicts logits $s_l$ against level-$l$ targets $m_l$ and $v_l$ ($l=0$ full resolution; $l>0$ obtained by adaptive average pooling of $m$ and $v$ followed by binarization).
$\mathcal{L}_{\mathrm{seg}}$ is the uniform sum over four scales of masked binary cross-entropy plus soft Dice,
\begin{equation}
  \mathcal{L}_{\mathrm{seg}} = \sum_{l=0}^{3}
    \Big[\, \mathrm{BCE}(s_l, m_l; v_l) + \mathrm{Dice}(s_l, m_l; v_l) \Big].
  \label{eq:lseg}
\end{equation}

For external task guidance, let $\mathcal U$ be the released post-contrast nnU-Net~\cite{isensee2021nnunet} from MAMA-SYNTH~\cite{joship2026mamasynth}, and let $\mathcal N(\cdot;v)$ denote per-image $z$-scoring over valid pixels $v$.
With tumor logits $z=\mathcal U(\mathcal N(\hat y,v))$,
\begin{equation}
    \mathcal{L}_{\mathrm{dseg}}= 0.4\,\mathcal{L}_{\mathrm{BCE}}(z,m;v)+0.6\,\mathcal{L}_{\mathrm{Dice}}(z,m;v).
\end{equation}
$\mathcal U$ remains frozen, but gradients flow through $\hat y$.
Training applies a single full-resolution forward pass on the padded canvas, without the evaluator's sliding-window inference, test-time mirroring, or full nnU-Net preprocessing pipeline; $\mathcal{L}_{\mathrm{dseg}}$ is therefore a differentiable surrogate of, not an exact reproduction of, the deployed segmentation score.

\subsection{Optimization, baselines, and evaluation}
MIRAGE is trained with Adam, initial learning rate $5\!\times\!10^{-4}$, warm-up cosine decay, batch size 4, and 59,000 iterations on a single NVIDIA H100 GPU. 
% Checkpoints are selected based on validation. 
% A later final refit, not used for the reported comparisons, ran for 71,000 iterations on a 97/3 split.

% 59k iters / 1205 samples  = 49 epochs

% \subsection{Baselines}

We compare with tuned paired-translation implementations of pix2pix~\cite{isola2017pix2pix}, a conditional DDPM~\cite{ho2020ddpm} predicting the subtraction image, and one-step LBM~\cite{chadebec2025lbm}. 
% All use the same settings. 
% The experiments compare concrete tuned systems; they do not establish exhaustive optimization of each paradigm.

% \subsection{Evaluation}

All models are evaluated on the validation cohort using the eight metrics defined by the MAMA-SYNTH protocol~\cite{joship2026mamasynth}. The metrics cover four complementary aspects of synthesis quality. Global image fidelity is measured with MSE and LPIPS~\cite{zhang2018lpips}, while tumor-region realism is assessed with tumor-masked SSIM~\cite{wang2004ssim} and Fr\'echet radiomic distance (FRD), which compares radiomic-feature distributions extracted from synthetic and reference tumor regions~\cite{konz2026frechet}.

Two fixed radiomics classifiers provide task-oriented measures. AUROC$_\mathrm{C}$ evaluates whether the synthetic images exhibit a post-contrast enhancement phenotype by measuring their separability from pre-contrast images. AUROC$_\mathrm{R}$ distinguishes the annotated tumor ROI from its mirrored contralateral counterpart and hence probes whether tumor-associated enhancement is present at the correct region. Both classifiers use the reference tumor mask and consequently assess enhancement characteristics rather than automatic tumor localization.

Downstream utility is evaluated with a released by MAMA-SYNTH single-fold 2D nnU-Net. Predictions on the synthetic images are compared with the reference tumor masks using Dice and the $95$th-percentile Hausdorff distance (HD95). This evaluation uses the complete released nnU-Net preprocessing and inference pipeline, in contrast to the simplified differentiable forward pass used to compute $\mathcal{L}_{\mathrm{dseg}}$ during training.

For experiments repeated across random seeds, results are reported as mean $\pm$ std. Differences between configurations are assessed using the two-sample standard error

\begin{equation}
    \mathrm{SE}_{\Delta} = \sqrt{{\sigma_1^2}/{n_1} + {\sigma_2^2}/{n_2}},
\end{equation}
rather than the variability of either configuration alone. A difference is marked when its magnitude is at least $2,\mathrm{SE}_{\Delta}$.

% The protocol reports eight metrics in four groups~\cite{joship2026mamasynth}: global MSE and LPIPS; tumor-region SSIM and Fr\'echet radiomic distance (FRD), which compares radiomic-feature distributions~\cite{wang2004ssim,konz2026frechet}; two fixed radiomics-classifier AUROCs; and Dice/HD95 from the a released single-fold 2D nnU-Net \cite{joship2026mamasynth}. AUROC$_\mathrm{C}$ measures post- versus pre-contrast separability, whereas AUROC$_\mathrm{R}$ distinguishes the reference tumor ROI from its mirrored contralateral ROI. Both use the reference mask and therefore assess enhancement phenotype rather than automatic localization. 

% Ablations report mean $\pm$ standard deviation over three seeds. $^\ast$ denotes a mean difference exceeding twice the combined standard error relative to the stated reference; $^\dagger$ denotes complete seed separation without meeting that criterion. These are descriptive stability indicators, not multiplicity-corrected tests.

\section{Results}

\begin{figure}[!ht]
  \centering
  \includegraphics[width=1\textwidth]{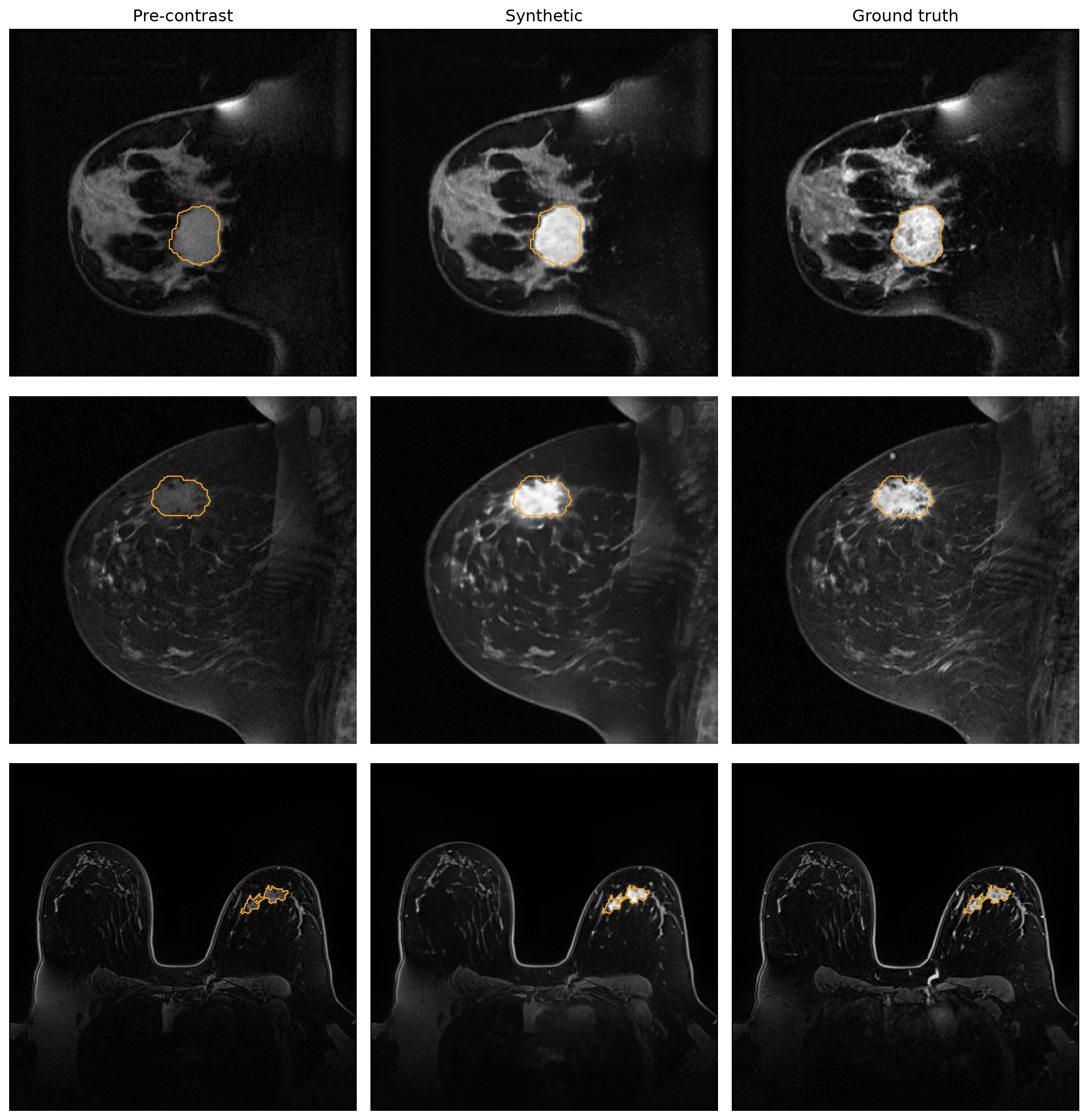}
  \caption{Qualitative synthesis on three held-out cases from different cohorts, each of single-breast view. The contour indicates the expert annotated tumor.}
  \label{fig:qualitative}
\end{figure}

Figure~\ref{fig:qualitative} shows representative results from different cohorts and acquisition orientations. MIRAGE preserves the overall breast anatomy and concentrates the predicted enhancement within the annotated tumor region, although fine differences in intratumor texture and surrounding parenchymal enhancement remain visible relative to the reference.

\subsection{Loss ablations}

The contribution of each auxiliary objective is examined from two complementary directions. The leave-one-in experiments in Table~\ref{tab:loi} add each term individually to the masked-$\Lone$ baseline, whereas the leave-one-out experiments in Table~\ref{tab:loo} remove one term at a time from the complete objective.

\begin{table}[h!]
  \caption{Leave-one-in ablation over three seeds. Each loss is added to the masked-$L_1$ baseline. $^{\ast}$: mean difference $>2$ combined standard errors relative to $L_1$.}
  \label{tab:loi}
  \centering
  \footnotesize
  \resizebox{\textwidth}{!}{%
    \renewcommand{\arraystretch}{1.25}%
    \begin{tabular}{@{}lcccccccc@{}}
      \toprule
      & \multicolumn{2}{c}{Image fidelity}
      & \multicolumn{2}{c}{ROI realism}
      & \multicolumn{2}{c}{Classification}
      & \multicolumn{2}{c}{Segmentation} \\
      \cmidrule(lr){2-3}
      \cmidrule(lr){4-5}
      \cmidrule(lr){6-7}
      \cmidrule(lr){8-9}
      Recipe
      & MSE\,\dn
      & LPIPS\,\dn
      & FRD\,\dn
      & SSIM$_\text{tum}$\,\up
      & AUROC$_\text{C}$\,\up
      & AUROC$_\text{R}$\,\up
      & Dice\,\up
      & HD95\,\dn \\
      \midrule
      \textbf{$L_1$ only}
      & $0.597_{\pm 0.009}$
      & $0.173_{\pm 0.003}$
      & $10.47_{\pm 0.36}$
      & $0.482_{\pm 0.017}$
      & $0.721_{\pm 0.009}$
      & $0.683_{\pm 0.005}$
      & $0.495_{\pm 0.038}$
      & $121.7_{\pm 17.6}$ \\

      $+\,\Llpips$
      & $0.603_{\pm 0.004}$
      & $0.091_{\pm 0.002}^{\ast}$
      & $7.23_{\pm 0.12}^{\ast}$
      & $0.498_{\pm 0.013}$
      & $0.726_{\pm 0.006}$
      & $0.694_{\pm 0.007}^{\ast}$
      & $0.562_{\pm 0.007}^{\ast}$
      & $98.7_{\pm 3.8}^{\ast}$ \\

      $+\,\Lseg$
      & $0.568_{\pm 0.009}^{\ast}$
      & $0.172_{\pm 0.005}$
      & $10.50_{\pm 1.47}$
      & $0.543_{\pm 0.018}^{\ast}$
      & $0.660_{\pm 0.006}^{\ast}$
      & $0.712_{\pm 0.003}^{\ast}$
      & $0.635_{\pm 0.010}^{\ast}$
      & $56.5_{\pm 7.6}^{\ast}$ \\

      $+\,\Ldseg$
      & $0.596_{\pm 0.011}$
      & $0.164_{\pm 0.005}^{\ast}$
      & $7.77_{\pm 0.64}^{\ast}$
      & $0.511_{\pm 0.008}^{\ast}$
      & $0.795_{\pm 0.002}^{\ast}$
      & $0.681_{\pm 0.007}$
      & $0.634_{\pm 0.016}^{\ast}$
      & $58.1_{\pm 10.8}^{\ast}$ \\

      $+\,\Lunder$
      & $0.599_{\pm 0.004}$
      & $0.170_{\pm 0.002}$
      & $9.81_{\pm 0.83}$
      & $0.554_{\pm 0.004}^{\ast}$
      & $0.698_{\pm 0.009}^{\ast}$
      & $0.713_{\pm 0.016}^{\ast}$
      & $0.632_{\pm 0.025}^{\ast}$
      & $54.9_{\pm 11.3}^{\ast}$ \\
      \bottomrule
    \end{tabular}%
  }
\end{table}

\begin{table}[t]
  \caption{Leave-one-out ablation over three seeds. Each loss is removed from the full objective. $^{\ast}$: mean difference $>2$ combined standard errors; $^{\dagger}$: complete seed separation without meeting that criterion.}
  \label{tab:loo}
  \centering
  \footnotesize
  \resizebox{\textwidth}{!}{%
    \renewcommand{\arraystretch}{1.25}%
    \begin{tabular}{@{}lcccccccc@{}}
      \toprule
      & \multicolumn{2}{c}{Image fidelity}
      & \multicolumn{2}{c}{ROI realism}
      & \multicolumn{2}{c}{Classification}
      & \multicolumn{2}{c}{Segmentation} \\
      \cmidrule(lr){2-3}
      \cmidrule(lr){4-5}
      \cmidrule(lr){6-7}
      \cmidrule(lr){8-9}
      Recipe
      & MSE\,\dn
      & LPIPS\,\dn
      & FRD\,\dn
      & SSIM$_\text{tum}$\,\up
      & AUROC$_\text{C}$\,\up
      & AUROC$_\text{R}$\,\up
      & Dice\,\up
      & HD95\,\dn \\
      \midrule
      \textbf{Full}
      & $0.607_{\pm 0.008}$
      & $0.095_{\pm 0.001}$
      & $6.56_{\pm 0.15}$
      & $0.540_{\pm 0.007}$
      & $0.711_{\pm 0.031}$
      & $0.718_{\pm 0.013}$
      & $0.679_{\pm 0.009}$
      & $34.8_{\pm 2.3}$ \\

      $-\,\Ldseg$
      & $0.590_{\pm 0.010}^{\ast}$
      & $0.093_{\pm 0.001}^{\ast}$
      & $6.98_{\pm 0.22}^{\ast}$
      & $0.553_{\pm 0.004}^{\ast}$
      & $0.712_{\pm 0.011}$
      & $0.712_{\pm 0.013}$
      & $0.667_{\pm 0.010}$
      & $45.5_{\pm 4.1}^{\ast}$ \\

      $-\,\Lseg$
      & $0.601_{\pm 0.013}$
      & $0.095_{\pm 0.001}$
      & $6.78_{\pm 0.17}$
      & $0.543_{\pm 0.005}$
      & $0.758_{\pm 0.035}$
      & $0.708_{\pm 0.003}$
      & $0.663_{\pm 0.004}^{\ast}$
      & $45.2_{\pm 3.1}^{\ast}$ \\

      $-\,\Llpips$
      & $0.608_{\pm 0.009}$
      & $0.167_{\pm 0.001}^{\ast}$
      & $8.74_{\pm 2.13}^{\dagger}$
      & $0.521_{\pm 0.010}^{\ast}$
      & $0.732_{\pm 0.008}$
      & $0.735_{\pm 0.009}^{\dagger}$
      & $0.671_{\pm 0.002}^{\dagger}$
      & $36.1_{\pm 3.0}$ \\

      $-\,\Lunder$
      & $0.581_{\pm 0.004}^{\ast}$
      & $0.094_{\pm 0.002}$
      & $6.79_{\pm 0.51}$
      & $0.530_{\pm 0.005}^{\ast}$
      & $0.718_{\pm 0.009}$
      & $0.717_{\pm 0.011}$
      & $0.674_{\pm 0.001}$
      & $34.4_{\pm 2.5}$ \\
      \bottomrule
    \end{tabular}%
  }
\end{table}

The masked-$\Lone$ model performs poorly on downstream segmentation. Adding any of the four auxiliary terms significantly improves both Dice and HD95. The three mask-aware terms reach Dice scores within $0.003$ of one another, showing that each can independently recover a large part of the localization performance.

The terms differ more clearly outside the segmentation metrics. Adding either $\Llpips$ or $\Lunder$ improves image fidelity, tumor-masked SSIM, and radiomic agreement. Adding $\Lseg$ or $\Ldseg$ produces the largest gains in tumor localization and also improves tumor-region realism.

The complete objective achieves better segmentation performance than any leave-one-in configuration. In the leave-one-out analysis, removing $\Lseg$ is the only ablation that significantly reduces Dice. Removing $\Ldseg$ significantly worsens both FRD and HD95, while Dice remains essentially unchanged. Removing $\Llpips$ worsens LPIPS and also increases FRD.

The effect of $\Lunder$ differs between the two analyses. It provides clear improvements when added alone to the masked-$\Lone$ model, but removing it does not materially alter FRD, Dice, or HD95. Its remaining effect is a small improvement in SSIM$_\mathrm{tum}$ accompanied by a small increase in MSE.

All four leave-one-out configurations produce a higher mean AUROC$_\mathrm{C}$ than the complete objective. None of these differences reaches the $2\sigma$ criterion, however, and the complete model has the largest variability for this metric, with a standard deviation of $\pm0.031$.

\subsection{Comparison across synthesis paradigms}
\label{sec:paradigms}

\begin{table}[!t]
  \caption{Cross-paradigm comparison on the validation cohort. Average rank is the mean of the eight per-metric ranks and is descriptive rather than the benchmark's group-wise ranking.}
  \label{tab:paradigms}
  \centering
  \small
  \resizebox{\textwidth}{!}{%
    \begin{tabular}{@{}lccccccccc@{}}
      \toprule
      & \multicolumn{2}{c}{Image fidelity}
      & \multicolumn{2}{c}{ROI realism}
      & \multicolumn{2}{c}{Classification}
      & \multicolumn{2}{c}{Segmentation}
      & \\
      \cmidrule(lr){2-3}
      \cmidrule(lr){4-5}
      \cmidrule(lr){6-7}
      \cmidrule(lr){8-9}
      Model
      & MSE\,\dn
      & LPIPS\,\dn
      & FRD\,\dn
      & SSIM$_\text{tum}$\,\up
      & AUROC$_\text{C}$\,\up
      & AUROC$_\text{R}$\,\up
      & Dice\,\up
      & HD95\,\dn
      & Avg rank\,\dn \\
      \midrule
      MIRAGE (ours)
      & \textbf{0.607}
      & 0.095
      & \textbf{6.56}
      & \textbf{0.540}
      & 0.711
      & \textbf{0.718}
      & \textbf{0.679}
      & \textbf{34.8}
      & \textbf{1.50} \\
      DDPM
      & 0.795
      & 0.122
      & 9.07
      & 0.389
      & \textbf{0.816}
      & 0.632
      & 0.495
      & 122.4
      & 3.13 \\
      pix2pix
      & 0.628
      & 0.169
      & 11.75
      & 0.497
      & 0.755
      & 0.677
      & 0.509
      & 106.2
      & 2.88 \\
      LBM
      & 0.612
      & \textbf{0.079}
      & 8.12
      & 0.529
      & 0.792
      & 0.657
      & 0.493
      & 126.6
      & 2.50 \\
      \bottomrule
    \end{tabular}%
  }
\end{table}

Table~\ref{tab:paradigms} compares MIRAGE with pix2pix, conditional DDPM, and LBM on the same validation cohort and evaluation pipeline. The methods occupy different operating points across the metric groups. LBM achieves the best LPIPS, while DDPM obtains the highest AUROC$_\mathrm{C}$. MIRAGE achieves the best downstream segmentation performance and the lowest FRD. All three alternative synthesis models perform worse than MIRAGE on Dice and HD95. They also produce higher FRD values, despite obtaining stronger results on selected appearance-oriented metrics. Thus, the models producing the most perceptually favorable or strongly contrast-like outputs do not produce the most accurate tumor localization or the closest tumor-region radiomic distribution.

\section{Discussion and Conclusion}

The ablations show that the auxiliary objectives provide overlapping but distinct constraints. All improve the masked-$\Lone$ baseline individually, while the complete objective achieves better downstream segmentation than any single-term addition. The final performance therefore cannot be attributed to one dominant loss.

The segmentation objectives are particularly complementary. Removing $\Lseg$ is the only ablation that significantly reduces Dice, indicating that direct multiscale supervision remains important for overlap-based localization. Removing $\Ldseg$, by contrast, primarily worsens HD95 and FRD while leaving Dice largely unchanged, suggesting that guidance from the frozen segmenter additionally constrains lesion boundaries and tumor-region appearance. Removing $\Llpips$ likewise worsens both LPIPS and FRD. This association does not imply that LPIPS directly optimizes radiomic fidelity, but shows that perceptual and radiomic agreement are coupled in the present setting.

The clearest redundancy concerns $\Lunder$. Although effective when added alone, it has little effect on FRD, Dice, or HD95 once the other losses are present. Its remaining improvement in SSIM$_\mathrm{tum}$ is accompanied by a small increase in MSE. The other lesion-aware objectives may therefore already provide much of the pressure needed to preserve tumor enhancement, making $\Lunder$ the first candidate for removal when favoring a simpler objective.

The classification and cross-paradigm results reveal a broader trade-off. Removing any auxiliary term slightly increases AUROC$_\mathrm{C}$, although none of these differences reaches the predefined $2\sigma$ threshold. Similarly, LBM achieves the best LPIPS and DDPM the highest AUROC$_\mathrm{C}$, yet pix2pix, DDPM, and LBM all perform worse than MIRAGE on segmentation and FRD. Images that appear sharper or more characteristically post-contrast are therefore not necessarily better aligned with the patient-specific tumor location or radiomic distribution. The fixed classifier may reward a more stereotyped enhancement phenotype, whereas the lesion-aware losses emphasize spatial correspondence, although the observed AUROC$_\mathrm{C}$ trend remains tentative.

These differences are consistent with the respective objectives. Adversarial and generative distribution-learning methods encourage outputs that resemble the post-contrast domain, but realistic appearance alone does not guarantee correctly localised enhancement. 
The results concern the evaluated implementations and should not be interpreted as showing that adversarial, diffusion, or bridge-matching methods are intrinsically unsuitable.

The selected operating point is nevertheless conditional. Loss weights were chosen through coarse development sweeps on one internal validation cohort rather than exhaustive multi-objective optimisation, and different applications may prioritise different compromises between sharpness, localization, and downstream utility. Moreover, changes to the downstream model for $\mathcal{L}_{\text{dseg}}$ computation could alter the optimal objective and weighting.

Finally, none of the reported metrics establishes physiological correctness. Pixelwise metrics compare against one observed post-contrast acquisition, radiomic metrics capture selected image properties, and fixed classifiers and segmenters inherit their own biases. External cohorts, multiple acquisition protocols and downstream models, and reader studies are therefore needed to determine whether synthetic enhancement preserves diagnostically relevant findings without introducing, displacing, or suppressing lesions.

In conclusion, MIRAGE synthesizes peak post-contrast breast MRI from a single pre-contrast slice using tumor-informed supervision available only during training. It provides a stronger overall balance of patient-specific fidelity, tumor-region radiomic agreement, and downstream segmentation than the evaluated generative baselines, while conceding selected appearance-oriented metrics. More broadly, the findings support task-aware medical image synthesis in which pretrained downstream models guide outputs towards preserving clinically relevant information, rather than merely plausible appearance.

\clearpage
\bibliography{refs}

\bibliographystyle{unsrt}

\end{document}